\documentstyle[natbib209,aasj,mnfixes,epsfig]{mn-nat}   
\begin{document}

\title[Crust-core coupling and r-mode damping]{Crust-core coupling and
r-mode damping in neutron stars: a toy model}

\author[Levin \& Ushomirsky]
{Yuri Levin$^1$ and Greg Ushomirsky$^2$ \\
$^1$Astronomy Department, 601 Campbell Hall, University of
California, Berkeley, California, 94720;\\ yurlev@fire.berkeley.edu\\
$^2$Theoretical Astrophysics, M/C 130-33, Caltech, Pasadena,
California, 91125;\\ gregus@tapir.caltech.edu}
\date{June 1, 2000}

\maketitle

\begin{abstract}
R-modes in neutron stars with crusts are damped by viscous friction at
the crust-core boundary. The magnitude of this damping, evaluated by
Bildsten and Ushomirsky (BU) under the assumption of a perfectly rigid
crust, sets the maximum spin frequency for a neutron star spun up by
accretion in a Low-Mass X-ray binary (LMXB).  In this paper we explore
the mechanical coupling between the core r-modes and the elastic
crust, using a toy model of a constant density neutron star with a
constant shear modulus crust.  We find that, at spin frequencies in
excess of $\approx50$~Hz, the r-modes strongly penetrate the crust.
This reduces the relative motion (slippage) between the crust and the
core compared to the rigid crust limit.  We therefore revise down, by
as much as a factor of $10^2-10^3$, the damping rate computed by BU,
significantly reducing the maximal possible spin frequency of neutron
star with a solid crust.  The dependence of the crust-core slippage on
the spin frequency is complicated, and is very sensitive to the
physical thickness of the crust.  If the crust is sufficiently thick,
the curve of the critical spin frequency for the onset of the r-mode
instability becomes multi-valued for some temperatures; this is
related to the avoided crossings between the r-mode and the
higher-order torsional modes in the crust.  The critical frequencies
are comparable to the observed spins of neutron stars in LMXBs and
millisecond pulsars.

\end{abstract}

\section{Introduction and basic argument}
\label{sec:introduction}

Gravitational radiation offers an exciting possibility to set an upper
limit on the spin frequencies of accretionally spun-up neutron stars
in close binaries \citep{Bildsten98:GWs}. The source for gravitational
radiation may be the mass quadrupole supported by static stresses
within the neutron star crust (\citealt{Bildsten98:GWs},
\citealt*{UCB99}), or the current quadrupole from an $l=m=2$ r-mode
which is driven by gravitational radiation reaction
(\citealt{andersson99:accreting_rmode}, \citealt{Bildsten98:GWs}; for
a review of the r-mode instability see, e.g., \citealt{friedman99};
for the discovery paper see \citealt{andersson98:_new_class}).  In
this paper we concentrate on the second option, namely that
accretional spin-up triggers the r-mode instability, and the resulting
spin-down torque from gravitational radiation reaction balances or
exceeds the accretion torque. The instability is triggered when the
r-mode growth rate due to gravitational radiation reaction (which is a
sharply increasing function of the stellar spin frequency) becomes
equal to the r-mode viscous damping rate.

The viscous damping timescale for r-modes was first computed by
\citet*{lindblom98:_gravit_radiat_instab}. They assumed that a neutron
star is a fluid ball with a polytropic equation of state, and that the
viscosity is set by neutron-neutron scattering in the nuclear
matter. Their viscous damping implied that, if the neutron star core
temperature $T\gtrsim10^8$~K, which is the case for the rapidly
accreting neutron stars in LMXBs \citep{brown98,Brown99}, then the
r-mode instability would be triggered at a stellar spin frequency just
above $100$~Hz. This is inconsistent with observations: most neutron
stars in LMXBs are believed to be spinning with frequencies around
$300$~Hz (see \citealt{klis99:_millis} for a review), and the
existence of two 1.6~ms radio pulsars means that rapidly rotating
neutron stars are somehow formed in spite of the r-mode instability
\citep{andersson99:accreting_rmode}. The r-mode damping rate would
have to be $\sim 250$ higher than that computed by
\citet{lindblom98:_gravit_radiat_instab} to allow for such high spins.

\citet{lindblom98:_gravit_radiat_instab} conjectured that, for the
range of temperatures relevant for LMXBs, mutual friction between the
neutron and proton superfluids will provide the missing
damping. However, tour-de-force calculations of
\citet{lindblom99:superfluid} showed that, except for $\sim2$\% of the
allowed parameter space, mutual friction gives only a modest increase
of the r-mode damping rate, insufficient to account for the observed
high spins of neutron stars.

\citet{BU99} (hereafter BU) argued that the crust plays an important
role in r-mode damping. In their picture, the r-mode lives in the
liquid core, and most of the dissipation occurs in a thin (a few cm)
viscous boundary layer between the oscillating core and the motionless
(in the rotating frame) solid crust.  The damping BU computed is a
factor of $\sim10^5$ higher than that of
\citet{lindblom98:_gravit_radiat_instab}; it implies that the r-modes
would not become unstable unless the spin frequency is close to
break-up, which is significantly higher than the observed LMXB spins.
\citet*{andersson2000} and \citet{Rieutord2000} have revised the BU
estimate, down and up respectively, by factors of order unity. Both of
these corrections are valid; the net result is that the BU estimate is
almost unchanged.  However, there is still a substantial correction
unaccounted for in the previous work.  In this paper, we show that,
for a considerable fraction of the parameter space, the boundary layer
damping rate is a factor of $10^2-10^3$ smaller than the BU estimate,
and hence the critical frequencies for the onset of the r-mode
instability are correspondingly lower, $\sim200-600$~Hz.

BU considered only the leading-order transverse motion pattern of the
r-modes and assumed that these motions do not penetrate into the
crust. They estimated the magnitude of the viscous coupling between
these transverse motions and the crust, and found it to be small.
Therefore, their picture is that of a liquid in a bucket -- the
amplitude of the relative motion (slippage) between the crust and the
core is just equal to the amplitude of the fluid motion at the
boundary.  However, small radial motions of the r-mode, neglected by
BU, are very effective at coupling the crust and the core. Above a
small critical frequency, r-modes do {\it strongly} penetrate into the
crust, and the crust and the core oscillate almost in unison with very
similar amplitudes.  The slippage $\delta u$ in this regime is only a
small fraction of the total r-mode motion, $u$, with typical values of
$\delta u/u\sim 0.05-0.1$, which results in a correction of a factor
of $(\delta u/u)^2$ to the BU viscous boundary layer damping rate.

The fact that the crust is not rigid is central to our story. In fact,
the crust is more like a gel: the shear modulus is much smaller than
the bulk modulus.  The $l=m=2$ torsional mode in a non-rotating crust,
which has the same angular displacement pattern as the corresponding
r-mode, has a frequency $f_{\rm cr}\simeq 50$~Hz
\citep{McDermott88}. This frequency is a few times lower than the
r-mode frequency in rapidly rotating stars, indicating that the
elastic restoring force is quite weak compared to the Coriolis force
which is responsible for the r-mode oscillations.  Therefore, at
sufficiently high spin frequencies, one would expect the crust to
oscillate more or less like a liquid, with elasticity only slightly
modifying the mode structure.  Indeed, if the shear modulus $\mu$ of
the crust were exactly zero, i.e., if the crust were completely fluid,
there would be no slippage $\delta u$ between the crust and the
core. Since $\mu$ is non-zero but small, one would expect the slippage
to be proportional to the ratio of the elastic restoring force to the
Coriolis force, i.e., roughly $\delta u/u\sim(f_{\rm cr}/f_{\rm
rmode})^2$.  However, our numerical work shows that the slippage is
not a monotonically decreasing function of the spin frequency; it
rises sharply at the frequencies of avoided crossings between the
r-mode and the higher-order torsional modes of the crust.
Figure~\ref{fig:dispersion} shows the slippage as a function of the
spin frequency computed for two neutron star models with different
crust thicknesses (10\% and 20\% of the stellar radius).

We begin in \S~\ref{sec:equations} by outlining the equations
governing the pulsations of constant density stars with fluid cores
and solid crusts and presenting the numerical results for the
dependence of the slippage $\delta u/u$ on the spin frequency.  In
\S~\ref{sec:stability-curves} we use the computed slippages to
construct the critical r-mode stability curves; see
Figure~\ref{fig:instab}.  The resulting curves are quite sensitive to
the thickness of the neutron star crust.  In particular, for a
sufficiently thick crust, the stability curve has a peculiar
multi-valued structure, which is a consequence of the non-monotonic
dependence of the relative slippage on the spin frequency.  The
stability curve determines the maximum spin of a neutron star as a
function of temperature; we find that it generally goes through the
region of observed spin frequencies of neutron stars in LMXBs and in
millisecond radio pulsars.  Broad implications of our results are
discussed in \S~\ref{sec:discussion}.

\section{Core-crust boundary slippage}
\label{sec:equations}

\subsection{Basic formalism}
\label{sec:basic-formalism}

Oscillation modes of a rotating neutron star with an elastic crust
were analyzed by \citet{LS96} (hereafter LS). Their analysis was
performed for the case of slow rotation, which was treated as a
perturbation to other restoring forces, and the oblateness of the
equilibrium configuration was neglected.  We broadly utilize their
general formalism, however we derive equations specialized to the
constant-density case.  For r-modes the correction to the
eigenfrequency due to the non-sphericity of the equilibrium
configuration is of the same order as the correction due to the radial
motion \citep{Provost81:rmodes}.  However, we expect that the
oblateness of the star will not change the {\it qualitative\/} picture
for the crust-core slippage: at high enough spin frequencies, the core
and the crust will move almost in unison.  The {\it quantitative\/}
effects of non-sphericity are not treated in this paper.

For our toy model, we assume that the outer part, $r_{\rm cr}<r<R$, of the
neutron star of radius $R$ possesses a small shear modulus $\mu(r)$,
such that $\mu(r)={\rm const}$.  The neutron star fluid is assumed to
be incompressible, has a constant density $\rho$, and is rotating with
angular frequency $\Omega$.
 
The general displacement $\vec{u}$ of a fluid element, in a frame rotating
with the neutron star, is given by (\citealt{Zahn66}, LS):
\begin{equation}
\vec{u}=\sum_l
 (u_l(r) Y_{lm} \hat{r}+v_l(r)\nabla Y_{lm}
	+w_l(r)\hat{r}\times\nabla Y_{lm})e^{i\omega t},
\label{u1}
\end{equation}
where $\omega$ is the mode frequency in the rotating frame, $m$ is the
fixed azimuthal number, and $\nabla Y_{lm}$ is the gradient of the
spherical harmonic evaluated on a unit sphere.  We are interested in
the $l=m$ r-mode. For a purely fluid star, to ${\cal
O}(\Omega/\Omega_{\rm k})^2$, where $\Omega_{\rm k}=(GM/R^3)^{1/2}$ is
the keplerian angular velocity, the displacement for this mode is
given by
\begin{equation}
\vec{u}=\left(w_l \hat{r}\times \nabla Y_{ll}+v_{l+1}\nabla Y_{l+1\ l}
        +u_{l+1}Y_{l+1\ l}\hat{r}\right)e^{i\omega t},
\label{u2}
\end{equation}
i.e., $v_{l+1}/w_l$ and $u_{l+1}/w_l$ are of order
$(\Omega/\Omega_{\rm k})^2$.  In the solid crust, the displacement
vector need not satisfy the ordering assumed in Eq.~(\ref{u2}).
However, at least for modes with $\omega$ close to $2\Omega/3$, the
Coriolis force is the dominant restoring force, and we expect
$v_{l+1}/w_l$ and $u_{l+1}/w_l$ to be small; this ordering is
confirmed by our numerical results (see
\S~\ref{sec:stability-curves}).

We now define dimensionless frequencies
$\tilde{\Omega}=\Omega/\Omega_{\rm k}$ and
$\tilde{\omega}=\omega/\Omega_{\rm k}$, as well as the dependent 
variables similar to those utilized by LS:
\begin{mathletters}\label{eq:var-definitions}
\begin{eqnarray}
z_1 &=& \frac{u_{l+1}}{r} \\
z_2 &=& 2 \frac{du_{l+1}}{dr} \\
z_3 &=& \frac{v_{l+1}}{r} \\
z_4 &=& \frac{dv_{l+1}}{dr} - \frac{v_{l+1}}{r} + \frac{u_{l+1}}{r} \\
z_5 &=& i\frac{w_l}{r} \\
z_6 &=& i\left(\frac{d w_l}{dr}-\frac{w_l}{r}\right).
\end{eqnarray}
\end{mathletters}
In the constant density case, $z_2$ is not a dynamical variable, as it
can be eliminated using the continuity equation,
$\nabla\cdot\vec{u}=0$, or
\begin{equation}
\frac{1}{2}z_2+2z_1-(l+1)(l+2)z_3=0.
\end{equation}
However, even in the $\rho={\rm const}$ case, the Eulerian pressure
perturbation is non-zero, and, to leading order, is given by $\delta
p=\delta p_{l+1}(r) Y_{l+1\ l}$.  We define the dimensionless pressure
perturbation, $\tilde{p}=\delta p_{l+1}/\rho g r$, where $g=GM(r)/r^2$
is the local gravitational acceleration.  Finally, we define the
dimensionless shear modulus, $\tilde{\mu}=\mu/\rho gr$.  Throughout
this paper, we use $\tilde{\mu}=1.5\times10^{-4} (R/r)^2$, chosen so
that the the frequency of the $l=m=2$ fundamental torsional crustal
mode is close to the $\approx50$~Hz value computed by
\citep{McDermott88} for a non-rotating neutron star; the shear wave
speed corresponding to this choice is
$c_t=(\mu/\rho)^{1/2}=1.7\times10^8$~cm~s$^{-1}$.

These variables allow us to write the equations of motion in the crust
as a system of first-order differential equations:
\begin{mathletters}\label{eq:solid-eqns}
\begin{eqnarray}
r\frac{dz_1}{dr} &=& (l+1)(l+2) z_3 - 3 z_1 \\
r\frac{d\tilde{p}}{dr} &=& -2\tilde{p} + \tilde{\omega}^2 z_1-2\tilde{\omega}
                 \tilde{\Omega} l z_3 
                -2\tilde{\omega}\tilde{\Omega} l J_{l+1}^l z_5 \\ \nonumber
                &+&\tilde{\mu}(l+1)(l+2)(z_4+2z_3-2z_1) \\
r\frac{dz_3}{dr} &=& z_4-z_1 \\
r\frac{dz_4}{dr} &=& -2z_1+2\left[l(l+3)+1\right]z_3-3z_4 \\ \nonumber
 &+&
        \frac{1}{\tilde{\mu}}\left[\tilde{p}+\frac{2lJ_{l+1}^l
                \tilde{\omega}\tilde{\Omega}}{l+1}z_5 \right.\\
  \nonumber
      &+& \left.\frac{2l\tilde{\omega}\tilde{\Omega}}{(l+1)(l+2)}(z_1+z_3)
                -\tilde{\omega}^2 z_3\right] \\
r\frac{dz_5}{dr} &=& z_6 \\
r\frac{dz_6}{dr} &=& -3z_6 +\left\{l(l+1)-2\right\}z_5 \\ \nonumber
 &+&
        \frac{1}{\tilde{\mu}}\left[\left(\frac{2\tilde{\omega}
            \tilde{\Omega}}{l+1}-\tilde{\omega}^2\right)z_5
        +\frac{2J_{l+1}^l\tilde{\omega}\tilde{\Omega}}{l+1}
		\left(z_1+(l+2)z_3\right)\right]
\end{eqnarray}
\end{mathletters}
where $J_{l+1}^l=(2l+3)^{-1/2}$. The derivation of these equations is
analogous to that of Eqs.~(9)--(21) of LS.  In the liquid core, the
only nontrivial equations are
\begin{mathletters}\label{eq:liquid-eqns}
\begin{eqnarray}
r\frac{d\tilde{p}}{dr} &=& (l-1)\tilde{p} \\
r\frac{dz_1}{dr} &=& -(l+4)z_1 \\ \nonumber &-&
	\left(\tilde\omega-\frac{2\tilde\Omega}{l+1}\right)
	\frac{(l+1)^4 (2l+3)}{8l\tilde\Omega^3}\tilde{p},
\end{eqnarray}
\end{mathletters}
and 
\begin{equation}
z_5=-\frac{(l+1)}{2lJ_{l+1}^l}  \frac{\tilde p}{\tilde\omega\tilde\Omega}
\end{equation}
(see, e.g., \citealt{Kokkotas99} or LS).

Both at the crust-core boundary, and at the top of the crust, we
require the tangential traction to be zero, i.e.,
\begin{equation}
z_4=z_6=0.
\label{boundary1}
\end{equation}
We hence neglect the viscous coupling between the transverse r-mode
motions and the crust.  This is the coupling considered and found
negligible by BU.  At the crust-core boundary the radial displacement
$z_1$ is continuous, and so is the radial component of the traction,
$\Delta p - 2\mu du_{l+1}/dr$, or
\begin{mathletters}\label{boundary2}
\begin{eqnarray}
z_1(r_c^-) &=& z_1(r_c^+) \\ 
\tilde p(r_c^+) &-&\tilde{\mu}[2(l+1)(l+2)z_3(r_c^+)-4z_1]
=\tilde p(r_c^-).
\end{eqnarray}
\end{mathletters}
At the top of the crust, the radial component of the traction is zero:
\begin{equation}
\tilde p(R)-z_1(R)-\tilde{\mu}[2(l+1)(l+2)z_3(R)-4z_1(R)]=0
\label{boundary3}
\end{equation}
Equations (\ref{boundary1})---(\ref{boundary3}) provide the 
boundary conditions for the equations of motion.  We normalize our
solutions to have unit transverse displacement just inside the liquid
core,
\begin{equation}\label{eq:normalization}
z_5(r_c^-)=1.
\end{equation}
With this normalization, the slippage is $\delta
u/u=[(z_5(r_c^-)-z_5(r_c^+))^2+(z_3(r_c^-)-z_3(r_c^+))^2]^{1/2}$
(where we used our assumption, $z_3\ll z_5$).

\subsection{Numerical results}
\label{sec:numerical-results}

The keplerian spin frequency $\Omega_{\rm k}$, the shear modulus
$\mu$, and the ratio of the crust thickness to the stellar radius
$r_c/R$ are the only parameters necessary to describe a constant
density star (we neglect the thin surface ocean and the modes that are
present in it, as they are unlikely to influence the crust-core
slippage).  The crust thickness is roughly proportional to the scale
height at the crust-core boundary, and is hence $\propto R^2/M$. In
order to survey the range of parameters, we carry out our calculations
for two models, a ``thin'' crust model with $r_c/R=0.9$, and a
``thick'' crust model with $r_c/R=0.8$.  Because of the differences in
the normal mode spectrum between these two cases, we shall see that
the r-mode instability curves for the two models  have qualitatively
different behavior in the temperature range of interest for accreting
neutron stars, $T=10^8-10^9$~K.

\begin{figure*}
\hbox{
\epsfig{file=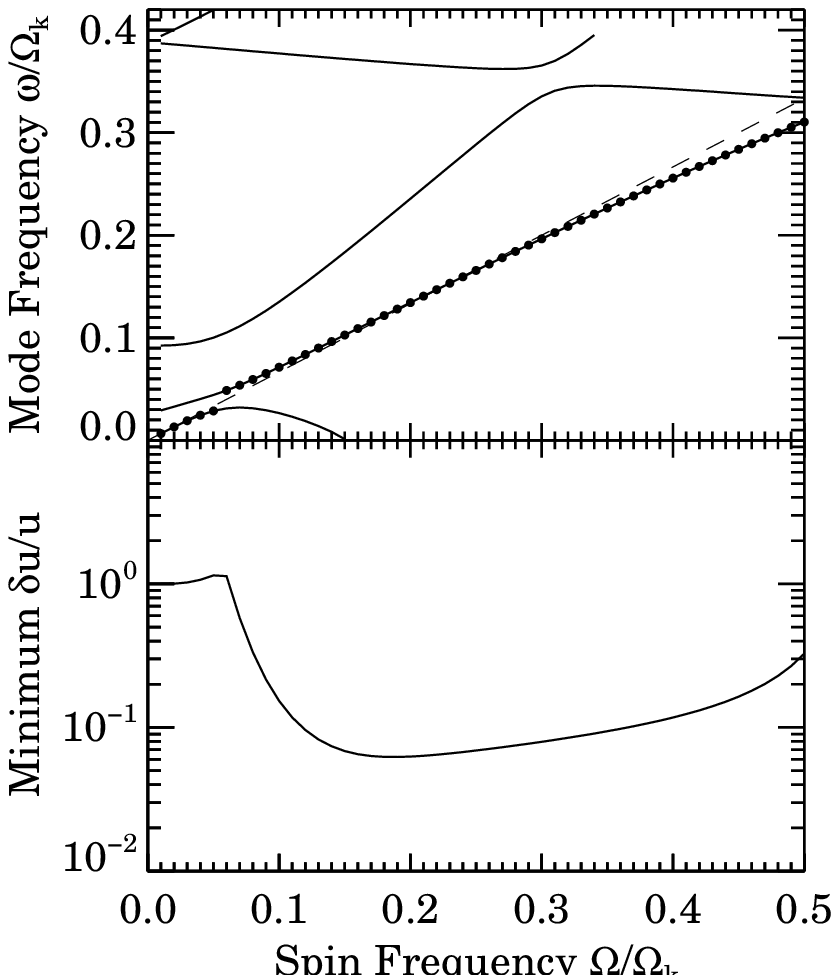}
\epsfig{file=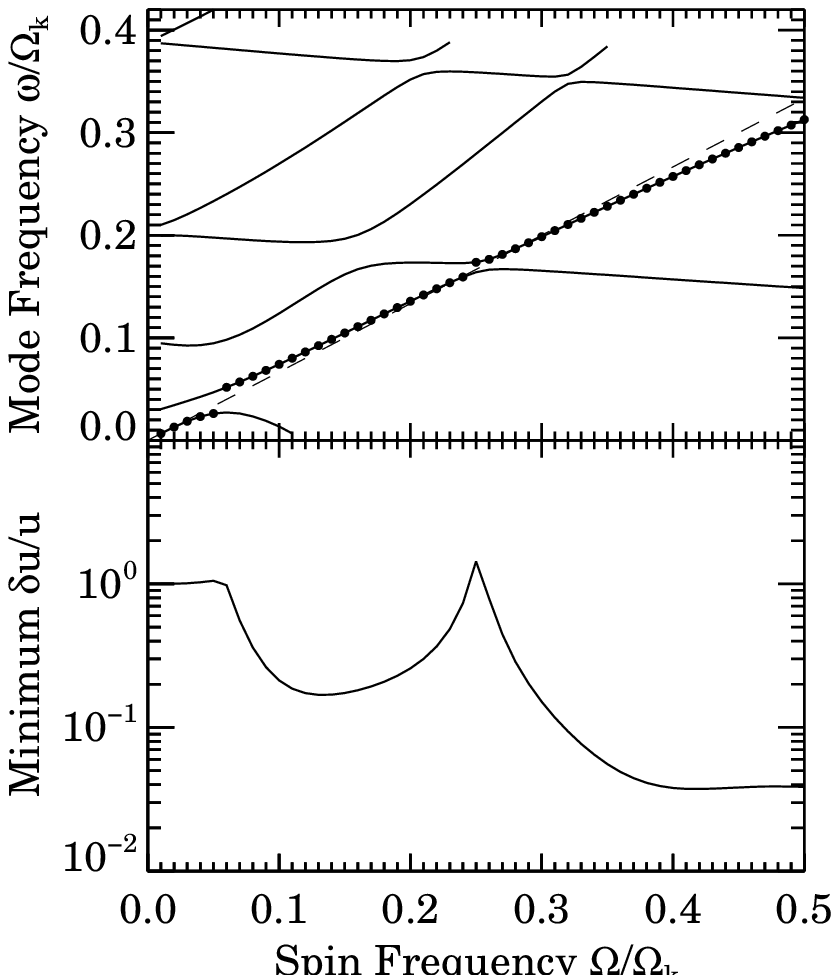}
}
\caption{\label{fig:dispersion} Top left: Mode angular frequency
$\omega$ (in the rotating frame) as a function of the angular spin
frequency $\Omega$ for the five lowest-frequency modes of the thin
crust model.  The dashed line shows the ``classical'' r-mode
dispersion relation, $\omega=2\Omega/3$, and the dots mark the mode
closest to this relation, which has the smallest slippage. Bottom
left: absolute value of the slippage of the mode marked with thick
dots in the top panel (i.e., the minimum slippage) as a function of
$\Omega$.  Right: same as top left, but for the thick crust model.}
\end{figure*}

The left panel of Figure~\ref{fig:dispersion} displays the results of
our calculations for $l=m=2$ for the thin crust model, while the right
panel shows the results for the thick crust model.  Let us first
concentrate on the top plots, which show the dispersion relation,
$\omega(\Omega)$, for the two models.  Contrary to the case of a
purely fluid star, where, in the isentropic case, there is only one
``classical'' $l=m$ r-mode, all the modes displayed in
Figure~\ref{fig:dispersion} have displacements of the form given by
Eq.~(\ref{u2}), with $w_l\propto r^l$ {\it in the core}.  However,
only the mode with $\omega$ closest to the ``classical'' r-mode value,
$2\Omega/3$, is of interest to us.  This mode is denoted by thick
dots in Figure~\ref{fig:dispersion}. When $\omega$ is much different
from $2\Omega/3$, the displacements in the crust are much bigger than
the displacements in the core, resulting in large slippage, and,
hence, large damping.  In essence, unless the mode frequency is close
to the r-mode value, the elasticity of the crust plays the dominant
role in determining the mode properties, and the mode is a crustal
mode, rather than a core r-mode.

Consider the low spin frequency ($\Omega\ll(\mu/\rho
R^2)^{1/2}\approx0.05\Omega_{\rm k}$) limit of our results.  In this
case, the mode with the lowest frequency in
Figure~\ref{fig:dispersion} is the ``classical'' $l=m$ r-mode, which
has the transverse displacement $w_2\propto r^2$ in the core and
negligible displacements in the crust.  This behavior (liquid in a
bucket) is the approximation used by BU.  The modes at higher
frequencies are torsional crustal oscillations, which have appreciable
displacements in the crust, and negligible displacements in the
core. The fundamental (nodeless) torsional mode has
$\omega=0.03\Omega_k$ for both models.  The lack of sensitivity of the
frequency of this mode to the thickness of the crust was first noted
by \citet{Hansen80}.  The modes at higher frequencies are overtones,
and we note that the mode spectrum for the thick crust model is denser
than that of the thin crust model.  Qualitatively, the mode frequency
is determined by fitting a certain number of nodes of the
eigenfunction between the boundaries of the crust, and hence a thicker
crust supports more modes.

As the spin frequency is increased beyond $\Omega\approx(\mu/\rho
R^2)^{1/2}$, the behavior of the modes changes.  The frequency of the
core r-modes rises to meet with the frequencies of the modes which
originally resided in the crust, resulting in avoided crossings.  At
the first avoided crossing, the core r-mode is coupled with the
lowest-order torsional mode of the crust; at this avoided crossing the
core r-mode strongly penetrates the crust. As the spin frequency
increases beyond the first crossing, the crust-core slippage decreases
along the branch closest to the ``classical'' r-mode frequency,
$\omega=2\Omega/3$ (marked by a dashed line in
Figure~\ref{fig:dispersion}).  Since the boundary layer damping rate
is proportional to $(\delta u/u)^2$, this branch is most interesting
for us as it corresponds to the r-mode with the lowest damping, and,
hence, the most unstable.

The relative slippage as a function of the spin frequency is plotted
in the bottom panels of Figure~\ref{fig:dispersion}. We always plot
slippage of the r-mode from the branch with the lowest damping, marked
by thick dots in Figure~\ref{fig:dispersion} (i.e., the most unstable
r-mode).  As the spin frequency increases beyond the first crossing,
the slippage decreases to $\approx0.1$.  However, the thick crust
model has another avoided crossing at $\Omega\approx0.25\Omega_{\rm
k}$, and the slippage rises at this frequency because the core r-mode
couples resonantly with a higher-order torsional crustal mode.  In the
thick crust model, the ``valley'' in which the second mode dominates
only extends to $\Omega\approx0.25\Omega_{\rm k}$. After the avoided
crossing, the third mode has the smallest slip,
$4\times10^{-2}\lesssim\delta u/u\lesssim 1$. Note that the slip value
is $\approx 1$ near the avoided crossing and is $\gtrsim0.1$ for
$0.25\Omega_{\rm k}\lesssim\Omega\lesssim0.3\Omega_{\rm k}$, or for
spin frequencies between $\sim500$ and $600$~Hz.  In this regime,
the viscous boundary layer damping rate approaches the high BU value, and
so the critical spin frequency for the r-mode instability is high
as well.  As we now show, the shape of the r-mode stability curve
depends quite sensitively on the thickness of the crust.

\section{Critical stability curves}
\label{sec:stability-curves}

The dissipation rate in the viscous boundary layer is set by the
amplitude of the relative motion between the core and the crust.  BU
assumed that the crust remains static in the rotating frame, and hence
the amount of boundary slippage $\delta u/u=1$.  However, as shown in
the previous section, the slippage $\delta u/u$ is generally
significantly less than $1$.  Therefore, the BU estimate of the
damping rate has to be multiplied by $(\delta u/u)^2$.  The corrected
r-mode damping rate is [cf.\ Eq.~(4) of BU]
\begin{equation}
{1\over \tau_{\rm vbl}}\simeq
0.01{\rm~s}^{-1}{R_6^2 F^{1/2}\over M_{1.4}T_8}{\rho_b\over \rho}\left({f
\over
\hbox{kHz}}\right)^{1/2}\left(\frac{\delta u}{u}\right)^2,
\label{damping}
\end{equation}
where $f=\Omega/2\pi$ is the neutron star spin frequency, $R_6$,
$M_{1.4}$ and $T_8$ are the radius, the mass and the core temperature
of the neutron star in units of $10$~km, $1.4 M_\odot$, and $10^8$~K,
respectively, $\rho$ is the density at the crust-core interface,
$\rho_b=1.5\times 10^{14}$~g~cm$^{-3}$ is the estimated value of this
density \citep{pethick95}, and $F$ is the fitting parameter used by
\citet{cutler87} to parameterize shear viscosity found by
\citet{flowers79}.  The Ekman layer correction introduced by
\citet{Rieutord2000} for the constant density case is of order unity
and is not included in Eq.~(\ref{damping}); see \citet{lou2000} for a
treatment applicable for more realistic, non-uniform density stars.
We also neglected the contribution of the elastic energy of the crust
to the total mode energy, as well as the small deviation of the mode
frequency from $\omega=2\Omega/3$.

The r-mode growth rate due to gravitational radiation is given by
\begin{equation}
{1\over \tau_{\rm gw}}\simeq 0.02{\rm~s}^{-1}M_{1.4}R_6^4\left({f\over
\hbox{kHz}} \right)^6.
\label{growth}
\end{equation}
Note that this expression is different from the one used by BU.
\citet{andersson2000} pointed out that BU incorrectly rescaled
$\tau_{\rm gw}$ used by \citet{lindblom98:_gravit_radiat_instab} to
the canonical $R=10$~km.  \citet{andersson2000} gave their own
estimate of the r-mode growth rate for a constant density
star. Eq.~(\ref{growth}) is the correct $\tau_{\rm gw}$ for an $n=1$
polytrope, neglecting the deviation of the transverse displacement
from $w_{l}\propto r^l$ in the crust, as well as the small deviation
of the mode frequency from $\omega=2\Omega/3$.

While we computed the slippage using a constant density model, we
utilize $\tau_{\rm vbl}$ (Eq.~[\ref{damping}]) and $\tau_{\rm gw}$
(Eq.~[\ref{growth}]) appropriate for a more realistic $n=1$
polytrope. Compared to more realistic models, using $\tau_{\rm gw}$
and $\tau_{|rm vbl}$ for a constant density model overestimates the
gravitational radiation growth rate (by putting more mass at larger
radii), and underestimates the strength of the boundary layer damping
(by virtue of a larger mode energy).  Both of these effects result in
lower critical frequencies for the onset of the r-mode instability.
Slippage calculations for more realistic models are in progress.

\begin{figure}
\begin{center}
\epsfig{file=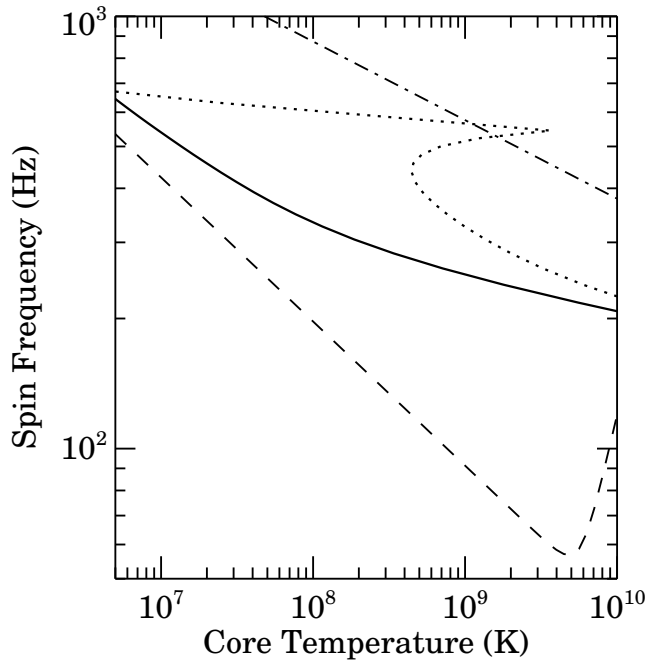}
\end{center}

\caption{\label{fig:instab} Critical frequencies as functions of the
core temperature for the modes in the thin crust model (solid line)
and the thick crust models (dotted line).  For comparison, we also
display the critical frequencies for the $n=1$ polytrope fluid model
(dashed line), and for a model with a crust and $\delta u/u=1$
independent of the spin (dot-dashed line).}
\end{figure}

The stability curve is obtained by equating the
gravitational-radiation-induced growth rate with the viscous
dissipation rate. Since we are primarily interested in temperatures
below the melting temperature of the crust, $\approx10^{10}$~K, we
neglect the dissipation due to bulk viscosity.  We also ignore the
dissipation due to the shear in the interior of the star, as well as
any potentially important dissipation mechanisms that may be operating
in the crust, and presume that the viscous boundary layer is the
dominant energy sink.  Since we do not have an analytical expression
for the slippage, it is more convenient to evaluate the critical
temperature, $T_8$, such that $\tau_{\rm gw}=\tau_{\rm vbl}$:
\begin{equation}
T_8=0.5 {F^{1/2}\over R_6^2 M_{1.4}^2}
{\rho_b\over\rho}
\left({f\over \hbox{kHz}}\right)^
{-5.5}\left({\delta u\over u}\right)^2.
\label{T8}
\end{equation}
Figure~\ref{fig:instab} shows the critical stability curves for models
with thin (solid line) and thick (dotted line) crusts, with $F$,
$M_{1.4}$, $R_6$ and $\rho_b/\rho$ all equal to one.  In other words,
the proper label for the abscissa of this plot is $T R_6^2 M_{1.4}^2
F^{-1/2} \rho/\rho_b$.  For comparison, we also plot the stability
curve for a purely fluid $n=1$ polytrope
(\citealt{lindblom98:_gravit_radiat_instab},
\citealt{Lindblom99:2ndorder}, dashed line), as well as the curve with
$\delta u/u=1$ independent of spin frequency (dash-dotted line,
cf. BU).  Neglecting the shear viscosity in the interior of the star
is clearly justified.

For the thin crust model (solid line), the slippage in the frequency
range of interest does not rise or fall sharply, and so the critical
frequency is a single-valued, monotonically decreasing function of
$T$.  However, for the neutron star with a thick crust, the slippage
and the damping rise sharply at the spin frequency close to the second
avoided crossing. Therefore, the critical temperature is not a
monotonically decreasing function of the spin frequency $f$, and the
stability curve is multi-valued.  Such a non-trivial stability curve
should be interpreted in terms of the critical temperature, rather
than the critical frequency: for temperatures larger than the critical
one given by Eq.~(\ref{T8}), i.e., for points in the $T$---$f$ plane
which lie to the right of the viscous boundary layer part of stability
curve, the r-mode is unstable and grows.  Otherwise, the r-mode is
stable and decays.

\section{Discussion, but no bets}
\label{sec:discussion}

Observationally, the spin frequencies of neutron stars in LMXBs have
been inferred from the QPO-like flux modulation during the type~I
X-ray bursts (the so-called burst QPOs, see \citealt{klis99:_millis}
for a review). In two out of the six sources displaying the burst
QPOs, the frequency is close to $300$~Hz, while in the remaining four
it is close to $600$~Hz.  In one of the latter four systems,
\citet{miller99:_eviden_antip_hot_spots_durin} found a subharmonic at
around $300$~Hz, and has argued that this subharmonic corresponds to
the true spin frequency.  Same relation is sometimes believed to hold
for the other $\approx600$~Hz bursters. In addition, $\approx300$~Hz
spin is often argued on the basis of a beat frequency interpretation
of the kHz QPO peak separation in all six of the burst QPO systems, as
well as for the neutron stars in thirteen additional LMXBs that do not
display burst QPOs.  On the other hand, there is no compelling
theoretical picture for why the subharmonics of the $\approx600$~Hz
burst QPOs should be so weak or invisible, so a bimodal spin
distribution is certainly not ruled out.

The new critical stability curves computed in this paper for two toy
models of neutron stars cut right through the observed range of spins
both for neutron stars in LMXBs and for millisecond pulsars, for
temperatures of interest for accreting neutron stars (between $10^8$~K
and $10^9$~K; \citealt{brown98}, \citealt{Brown99}).  An accreting
neutron star can continue its spin-up so long as it is located to the
left of the stability curve in Figure~\ref{fig:instab}. However, once
it crosses the critical stability curve and is located to the right
from it, the r-mode will grow and halt the spin-up, probably forcing
the neutron star into a thermal runaway cycle
\citep{levin99,spruit99:_gamma_x}.  The details of what happens during
this runaway and how violent it is are a topic of current research.
For example, the displacements induced in the crust may cause it break
when the strain (which is of order the r-mode amplitude $\alpha$, as
defined in \citealt{lindblom98:_gravit_radiat_instab}) exceeds a
critical value ($\lesssim10^{-2}$, see \S~6.1 of \citealt{UCB99} and
references therein for a summary of estimates of crustal breaking
strains) or melt if the heating due to the viscous boundary layer
raises its temperature above $\approx10^{10}$~K (see \citealt{lou2000}
for the calculation of the r-mode amplitude required to melt the
crust).  The evolution of the spin frequency and the temperature
depends on the saturation of the r-mode growth, which is not currently
understood.  Despite these uncertainties, it is clear that, if the
r-mode instability operates in accreting neutron stars, the
non-trivial shape of the stability curve and its sensitivity to the
crustal structure will be reflected in a peculiar, non-uniform spin
distribution of accreting neutron stars and millisecond pulsars.

Much work is necessary in order to make robust comparisons with
observations. R-mode calculations for more realistic neutron star
models, taking into account the complicated structure of the crust are
in progress.  We have not considered the possible superfluid nature of
neutrons and protons at the crust-core interface. In particular, we
have not considered the pinning of the superfluid vortices to the
crust. We assumed that the boundary layer is laminar, which may not be
true for large r-mode amplitudes.  Significant magnetic field at the
crust-core boundary may also affect the crust-core coupling
\citep{rezzolla2000,lai2000}. Thus, our estimate of the r-mode damping
may change when more complex physics is considered.

We thank Lars Bildsten for comments that helped improve this
manuscript, and Eliot Quataert and Re'em Sari for numerous
discussions.  YL thanks the Institute for Advanced Study, where part
of the research for this work was done, for hospitality.  YL is
supported by the Theoretical Astrophysics Center at UC Berkeley, and
GU is supported by NSF Grant AST-9618537 and NASA grant NGC5-7034.


\end{document}